# Tri-branched gels: Rubbery materials with the lowest branching factor approach the ideal elastic limit

Takeshi Fujiyabu[1], Naoyuki Sakumichi[1], Takuya Katashima[1], Chang Liu[2], Koichi Mayumi[2], Ung-il Chung[1], Takamasa Sakai[1]*

[1]*Department of Bioengineering, Graduate School of Engineering, The University of Tokyo, 7-3-1 Hongo, Bunkyo-ku, Tokyo, Japan.*
[2]*Material Innovation Research Center (MIRC) and Department of Advanced Materials Science, Graduate School of Frontier Sciences, The University of Tokyo, 5-1-5 Kashiwanoha, Kashiwa, Chiba 277-8561, Japan.*

*Corresponding author. Email: sakai@tetrapod.t.u-tokyo.ac.jp (T.S.)

**Abstract:**
Unlike hard materials such as metals and ceramics, rubbery materials can endure large deformations due to the large conformational degree of freedom of the crosslinked three-dimensional polymer network. However, the effect of the branching factor of the network on the ultimate mechanical properties of rubbery materials has not yet been clarified. This study shows that tri-branching, which entails the lowest branching factor, results in a large elastic deformation near the theoretical upper bound. This ideal elastic limit is realized by reversible strain-induced crystallization, providing on-demand reinforcement. The findings indicate that the polymer chain is highly orientated along the stretching axis, whereat enhanced reversible strain-induced crystallization is observed in the tri-branched and not in the tetra-branched network. A mathematical theory of structural rigidity is used to explain the difference in the chain orientation. Although tetra-branched polymers have been preferred since the development of vulcanization, these findings highlighting the merits of tri-branching will prompt a paradigm shift in the development of rubbery materials.

**Teaser:**
Tri-branched rubbery materials approach the ideal elastic limit via reversible strain-induced crystallization

**Main Text:**

**Introduction**

Vulcanization (1) is a chemical process that links polymers containing natural rubber and forms tetra-branched crosslinks, creating a three-dimensional (3D) polymer network in which the positions of the constituent polymer chains are fixed. Even when subjected to large deformations, the polymer chains return to their original position when the stress is removed (2). This is the molecular origin of the elasticity of rubber. Such rubbery materials have become essential in our daily lives, because their elasticity and flexibility provide a functionality range that cannot be supplied by hard materials. Despite the usefulness of these materials, they cause plastic pollution (3, 4) because such materials are a source of microplastic pollutants (5); the additives used in their preparation can also cause environmental pollution (6). Thus, there is a need to design new rubbery materials that are more environmentally friendly. A methodology for making rubbery materials stronger, without using any additives, should help meet this demand.

Although crosslinking is a well-known chemical process, its effect on the ultimate mechanical properties of rubbery materials is still not fully understood (7–10). One reason for the poor understanding is the invisibility of the polymer network: it is impossible to visualize the polymer network in rubber because the covalent bonds between the carbon atoms remain almost impossible to visualize using any state-of-the-art microscope. The crowding of the polymer chains also introduces challenges; polymer networks in rubbery materials are far removed from those shown schematically in Fig. 1A; instead, they resemble entangled spaghetti. The chain entanglements govern the mechanical properties of the rubbery materials and make it difficult to investigate the effects of the crosslinked structure on the properties (11).

One means of decreasing the number of chain entanglements is adding a diluent to the polymer network. The resultant swollen polymer network is called a polymer gel. In polymer gels, the polymer network architecture plays a vital role in determining the physical properties. Researchers have realized the advantages of such an approach in this century, and many polymer gels with advanced mechanical properties have been developed. These include gels with slidable crosslinks (12), gels consisting of two coexisting independent networks (13), homogeneous gels (14), and self-healing gels (15). Further, concepts developed in gel science have proven to be translatable, allowing the successful design and fabrication of excellent condensed rubbery materials (16–18). In this respect, gel science provides innovative methods for developing advanced rubbery materials.

This study focuses on elucidating the effect of the branching factor—the number of strands connected to each branch—on the ultimate mechanical properties of polymer networks. Previous studies have been mostly focused on tetra-branched networks, and systematic comparison with other branching factors has been limited to weak deformation regimes (19–21). In particular, little is known about tri-branching, which is the lowest branching factor that can form a network. It was found that tri-branching shows a large elastic deformation near the theoretical upper bound. This ideal elastic limit is realized by reversible strain-induced crystallization (SIC) providing on-demand reinforcement.

## Results

### Stretchability of tri-branched and tetra-branched gels

It is well known that the polymerization of mutually reactive tetra-branched polymeric precursors (i.e., prepolymers) in aqueous solution gives an ideally homogeneous polymer network, "soft diamond" (tetra-branched gel) (14) (Fig. 1A). The tetra-branched gel is also known as a model network for which the structure–property relationships are well understood (8, 14, 22-26). In this study, a homogeneous tri-branched polymer network was synthesized using a similar reaction with tri-branched prepolymers, resulting in a "soft triamond" (tri-branched gel). Schematics of both the tetra-branched and tri-branched hydrogels are shown in Fig. 1A. As shown in Fig. 1B, the tri-branched gel with a polymer concentration ($c$) of 140 g/L and the degree of polymerization of a strand connecting the branch points ($N$) of 610 was highly stretchable and did not break when stretched up to a stretch ratio ($\lambda$) of 30. The $\lambda$ is the maximum possible value that could be applied to an appropriate sample geometry in our apparatus. After stretching, the specimen lost 42% of volume due to water evaporation, and a strong residual strain was observed for the specimen (Fig. 1D), which mostly recovered over 24 hours (Fig. 1C and 1E). The recovery of the residual strain indicates that the residual strain is not caused by sample damage but by a strain-induced structure with slow dynamics. Notably, a tetra-branched gel with a similar composition ($c$ = 140 g/L and $N$ = 450) broke at $\lambda$ = 10, and little residual strain was observed (Fig. S1). This qualitative difference suggests that the tri-branched network endures the more extensive stretching compared to the tetra-branched one. In other words, upon subtracting a strand from each branch point, the stretchability is drastically enhanced.

To provide a systematic comparison, in addition to the tetra-branched and tri-branched gels, intermediate samples (tri-tetra-branched gels) were fabricated by reacting tri-branched and tetra-branched prepolymers with various concentration ratios (Fig. 1A). In the following, the mechanical properties of polymer gels in the as-prepared state are discussed. Figure 2A shows typical relationships between the stress ($\sigma$) and $\lambda$ of gels with different branching factors and $c$ = 140 g/L. Values of $c$ and $N$ were selected on the basis of an empirical scaling law of the stretch ratio at break ($\lambda_b$)

$$\lambda_b \sim c^{1/3} N^{2/3} \quad (1)$$

which was proposed previously for tetra-branched gels (22). This selection allowed us to safely stretch and break the specimens using our apparatus with negligible water evaporation. Under these conditions, all samples were elastic and did not show noticeable residual strains, even after breakage. Fig. 2B shows **Eq. 1** (dashed line in Fig. 2B) for the tetra-branched gels. **Equation 1** well reproduces $\lambda_b$ up to approximately 10 times the overlap concentration of prepolymers ($c^*$, Table 2), at which the polymer chains begin to overlap to fill the space, in the case of tetra-branched gels. In contrast, when the molar fraction of tri-branched prepolymers was gradually increased, $\lambda_b$ systematically deviated from **Eq. 1**, reaching a maximum for the purely tri-branched gel. The transition from the tetra-branched to tri-branched gel inevitably changes $N$ from 230 to 300 (see **Eq. 6** below), in addition to altering the branching factor, which can increase the $\lambda_b$ of the gel. However, the $N$-dependence of $\lambda_b$ in tetra-branched gels is well reproduced by **Eq. 1**. Further, the decreased elastic modulus of the tri-branched gel is not the origin of the increased $\lambda_b$ (see **Materials and Methods**), because such an enhancement was not observed in the tetra-branched gel with an elastic modulus ($G$) similar to that of the tri-branched gel (Fig. 2B). Therefore, the systematic deviation from **Eq. 1** suggests that the enhanced $\lambda_b$ of the tri-branched gel is a function of its branching factor.

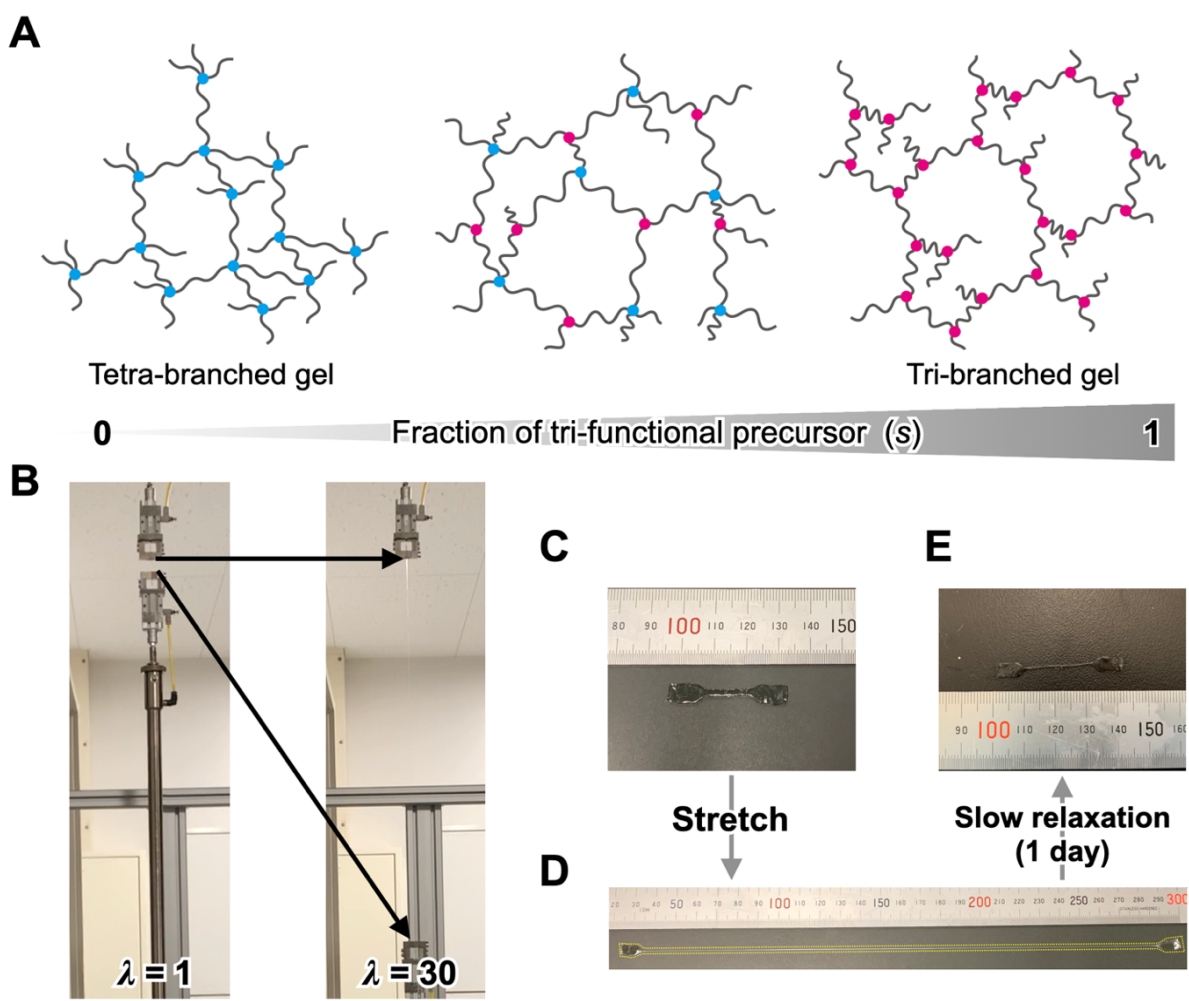

**Fig. 1. Schematics of target polymer networks and photographs of a stretching experiment.** (**A**) Schematics of a tetra-branched gel (left), tri-tetra-branched gel (middle), and tri-branched gel (right). These gels were systematically synthesized by mixing various ratios of tetra-branched and tri-branched prepolymers. (**B**) Photographs of a stretched tri-branched gel at stretch ratio $\lambda = 1$ and 30. Photographs of a dumbbell-shaped tri-branched gel sample (**C**) before and (**D**) after stretching ($\lambda = 30$). In (**D**), the gel is outlined by an overlaid dotted line as a guide to the eye. (**E**) When equilibrated at 25 °C without drying, the gel recovered its dumbbell shape. Photo Credit: Takeshi Fujiyabu, The University of Tokyo.

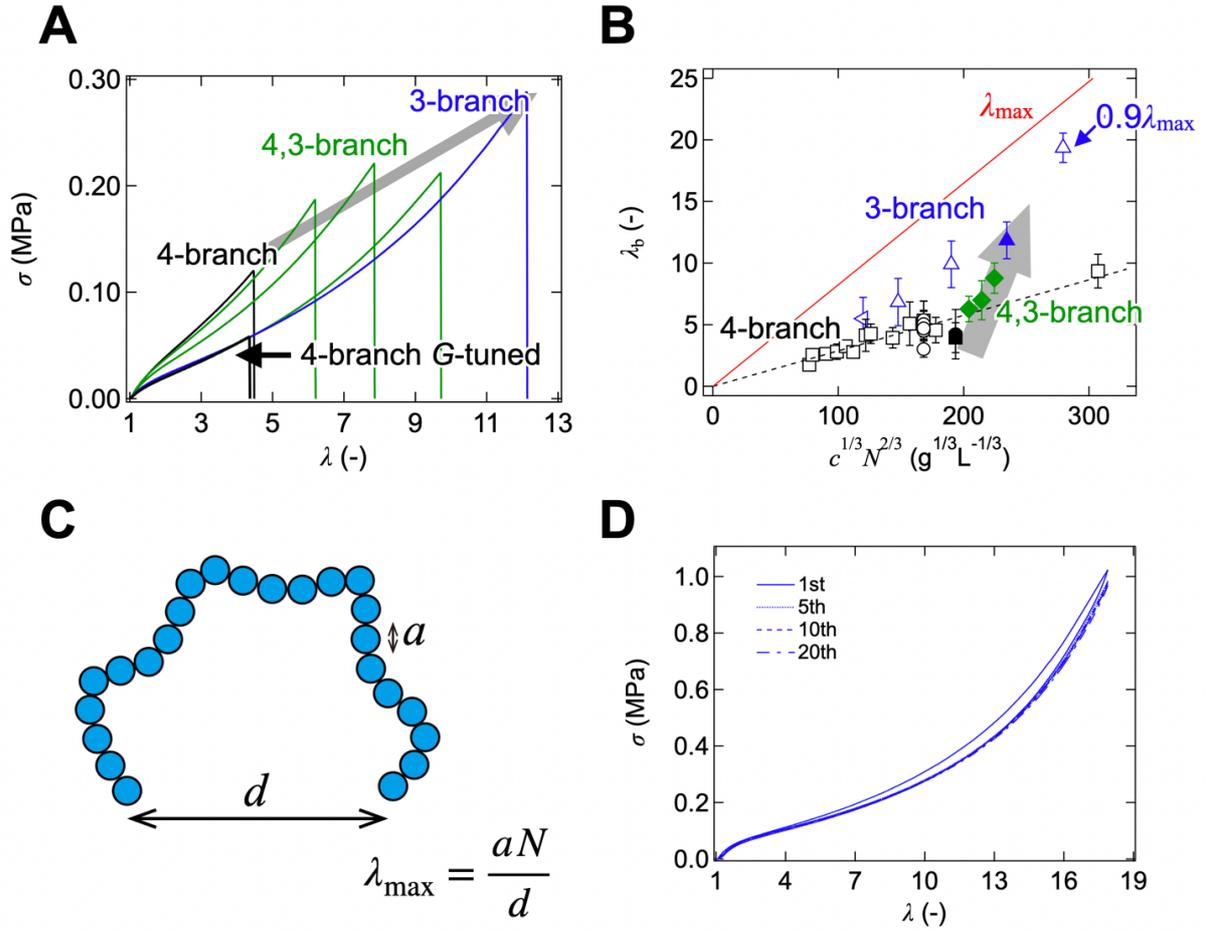

**Fig. 2. Mechanical properties of tri-branched, tri-tetra-branched, and tetra-branched gels.** (**A**) The $\sigma$–$\lambda$ relationships of a tri-branched gel (blue), tri-tetra-branched gels (green) with different molar fractions of tri-branched prepolymers, a tetra-branched gel (black), and a tetra-branched gel with the similar elastic modulus with the tri-branched gel (black). The $c$ was the same for all the gels (140 g/L). (**B**) $\lambda_b$ as a function of $c^{1/3}N^{2/3}$ for the tri-branched gels with different $c$ and $N$ (blue triangles), tri-tetra-branched gels with different $s$ defined in **Eq. 4** below (green rhombuses), tetra-branched gels with different $c$ and $N$ (black squares), and a tetra-branched gel with the similar elastic modulus with the tri-branched gel (black circles). Filled symbols correspond to the results shown in Fig. 2A. The values of $\lambda_b$ of the tetra-branched gels with $c < 140$ g/L were reproduced from our previous study (22). The black dashed line represents **Eq. 1** for the tetra-branched gels. The red solid line represents the maximum stretch ratio $\lambda_{max}$ estimated using **Eq. 2**. (**C**) The schematic picture showing the microscopic picture of maximum stretch ratio. (**D**) The $\sigma$–$\lambda$ relationships of tri-branched gels with $c = 240$ g/L under a cyclic loading test.

**Microscopic picture of the maximum stretch ratio**

We show that the empirical scaling law (**Eq. 1**) for the $\lambda_b$ of the tetra-branched gel is identical to a scaling law for the maximum stretch ratio $\lambda_{max}$ of a network strand. Here, $\lambda_{max}$ is defined as $\lambda_{max} \equiv aN/d$ (Fig. 2C). Here, $aN$ is the fully stretched length of a network strand and $a = 3.6$ Å is the effective length of a monomer unit (Fig. S2). Notably, the effect of crystallization on $\lambda_{max}$ was not considered, which agrees with the conventional treatment predicting the fully stretched length of a polymer chain (27). On the other hand, $d$ is the linear distance between the branch points of the polymer network. It is assumed that the average $d$ almost coincides with the distance between the centers of the adjacent prepolymers before the gelation reaction is initiated. The number density of prepolymers $n = cN_A/(m_{mono}N)$ satisfies $n \sim 1/d^3$, where $N_A$ is the Avogadro constant and $m_{mono} = 44$ g/mol is the molar mass of the monomer unit. Thus, using $n \sim c/N \sim 1/d^3$, we obtain $\lambda_{max} \sim N/d \sim c^{1/3}N^{2/3}$. The proportionality coefficient is denoted as $A$, and this result can be written as follows:

$$\lambda_{max} = Ac^{1/3}N^{2/3} \quad (2)$$

Here, $A$ can vary slightly depending on the placement of the branch point of the polymer network. Assuming that the center of the prepolymers (before the gelation reaction is initiated) has a random close-packed arrangement and that its placement determines the polymer network structure, then,

$$A = a\left(\frac{\pi N_A}{6\Phi m_{mono}}\right)^{1/3} \quad (3)$$

where $\Phi$ is the packing fraction of a random close packing, which varies slightly from very loose random packing ($\Phi \approx 0.56$) to close random packing ($\Phi \approx 0.63$ to $0.64$) of spheres (28). Here, $\Phi \approx 0.6$ is used, which is the value of the loose random packing of spheres. See **Materials and Methods** for the derivation of **Eq. 3**.

In Fig. 2B, $\lambda_{max}$ (red line in Fig. 2B) is plotted using **Eqs. 2** and **3**, indicating the upper bound of $\lambda_b$. Indeed, $\lambda_{max}$ is approximately three-fold the $\lambda_b$ of tetra-branched gels. This result agrees with the general rule for elastic materials: elastic materials do not endure the ideal strength or stretch ratio due to the structural heterogeneity, which induces heterogeneous stress concentration, crack formation, crack propagation, and subsequent macroscopic fracture (29). However, $\lambda_b$ approaches $\lambda_{max}$ when the molar fraction of the tri-branched prepolymers ($s$) increases, and eventually $\lambda_b \approx 0.6\lambda_{max}$ for tri-branched gels with $c = 140$ g/L (filled triangle in Fig. 2B). Because **Eqs. 2** and **3** predict the branching-factor-independent $\lambda_{max}$, the results indicate that the stress concentration is suppressed in the tri-branched gel. Notably, no existing models, including Kuhn's model for fracture (30), predict such a branching-factor-dependent $\lambda_{max}$.

There are two remarks on the above microscopic description predicting the maximum stretch ratio. First, $d$ is different from the end-to-end distance of a free polymer in the solution in Kuhn's model; the model is inappropriate to estimate $\lambda_{max}$ because it significantly underestimates $\lambda_{max}$ and predicts a weaker $c$-dependence of $\lambda_{max}$ than that observed experimentally. Further discussions of Kuhn's model are given in **Materials and Methods**.

Second, the microscopic description neglects any effect of trapped chain entanglements. This treatment is based on the successful correlation between the microscopic parameters and macroscopic mechanical properties, including elastic modulus (23, 24) and fracture energy (25), without considering any effects of trapped entanglements. Neglecting trapped entanglement is also justified based on the experimentally observed positive $c$-dependence of $\lambda_b$ in Fig. 2B (22). According to a model of entanglement (31), the effective chain length between neighboring entanglements ($N_{eff}$) decreases with increasing $c$ ($N_{eff} \sim c^{-1}$). When applying the scaling to our models, one obtains $\lambda_{max} \sim c^{-0.4}$, qualitatively contradicting the experimental results. Therefore, the effect of trapped entanglements is assumed to be negligible under the condition tested here.

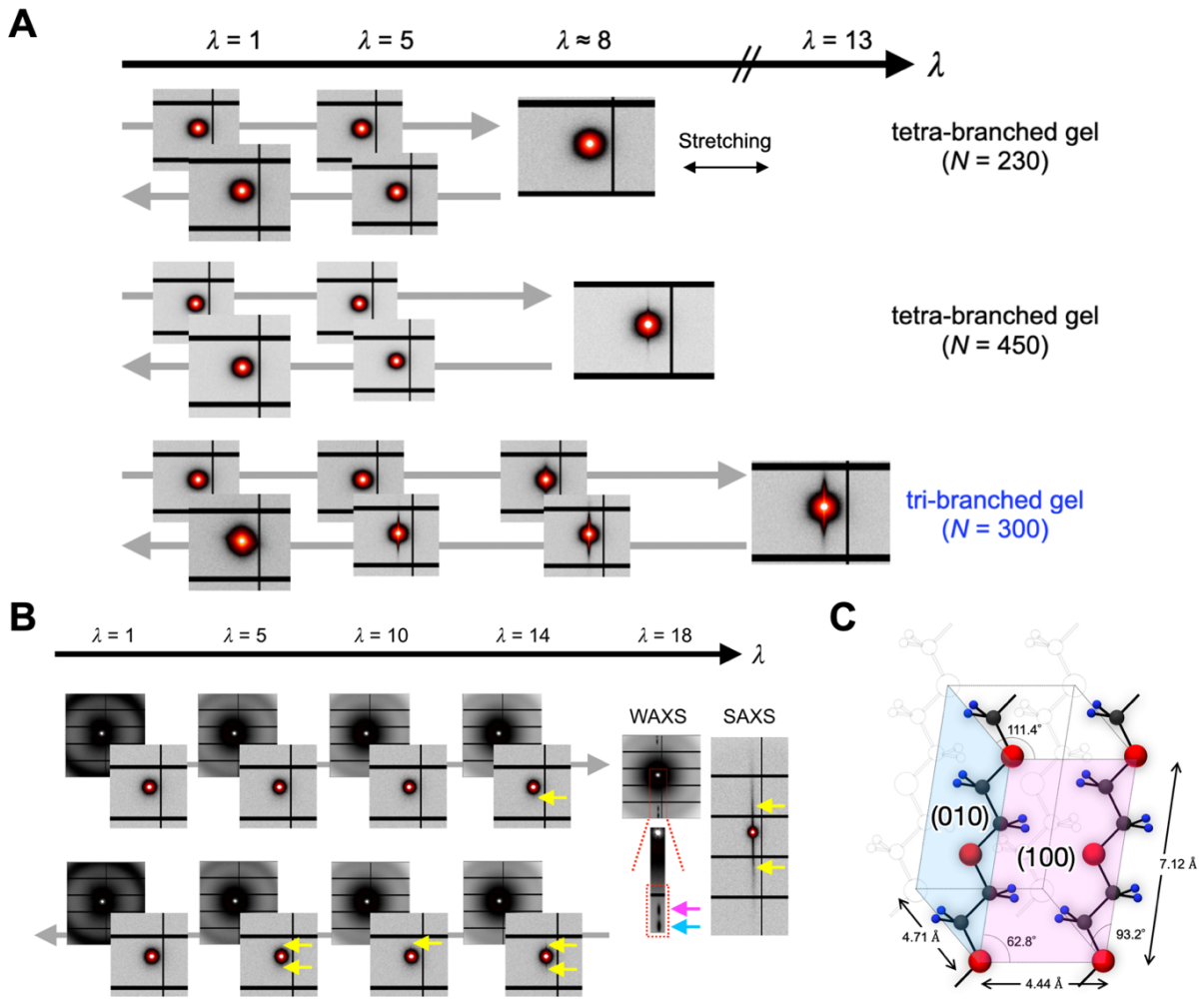

**Fig. 3. Results of SAXS/WAXS analyses of hydrogels during stretching.** (**A**) 2D SAXS patterns of tri-branched and tetra-branched gels with $c$ = 140 g/L acquired at various points during a loading–unloading cycle. (**B**) 2D SAXS/WAXS patterns of a 240 g/L tri-branched gel acquired in the range of $\lambda$ = 1.0–18 during a loading–unloading cycle. The red circles and yellow arrows have been overlaid to highlight diffraction spots and sharp streaks, respectively. The pink and blue arrows indicate (100) and (010) reflections of planar zig-zag PEG crystals, respectively. (**C**) Schematic of the planar zig-zag PEG crystal lattice with *trans* conformation.

**Strain-induced crystallization supporting ideal stretchability**

To investigate the structural changes that occurred during stretching, we performed small/wide angle X-ray scattering (SAXS/WAXS). Figure 3A shows the two-dimensional (2D) SAXS patterns of the tri-branched and tetra-branched gels acquired during a loading–unloading cycle; the samples were stretched close to $\lambda_b$ and then relaxed. Under the initial condition, only isotropic featureless patterns were observed, indicating that the tri-branched and tetra-branched gels had amorphous structures, unlike true triamond or diamond crystals schematically shown in Fig. 1A. Under moderate stretching conditions ($\lambda < 5$), all samples showed isotropic SAXS patterns, suggesting that the structure was isotropic (32). This is in contrast to conventional polymer gels, which show abnormal butterfly patterns under stretching due to structural heterogeneity (33). The absence of specific patterns in this region indicated good structural homogeneity for the tri-branched and tetra-branched gels. In contrast, under strong stretching conditions ($\lambda > 8$), a streak appeared in the SAXS patterns along the direction perpendicular to the stretching axis, suggesting the formation of strain-induced structure. Notably, such diluted polymer networks rarely show any strain-induced structure (34). Among the three sample types, the tri-branched gels produced the strongest SAXS pattern streak. This streak disappeared during the unloading process, indicating the reversibility of the strain-induced structure.

To further investigate the strain-induced structures of tri-branched gels, the polymer concentration was increased from 140 to 240 g/L. When the structural changes under stretching were investigated by SAXS/WAXS (Fig. 3B and S3, and Table 1), the tri-branched gel (240 g/L, $N = 300$) showed a sharp streak above $\lambda = 13$ and diffraction spots at $\lambda = 18$. The diffraction spots were indexed as (100) and (010) reflections of planar zig-zag crystals of the backbone polymer [poly(ethylene glycol) (PEG)] in the *trans* conformation (Fig. 3C) (35), suggesting that strain-induced crystallization (SIC) occurred. The high crystallinity, up to 80% (Table 1), indicates most network strands were fully stretched. Unexpectedly, $\lambda_b$ of the tri-branched gel almost corresponded to $\lambda_{max}$ ($\lambda_b \approx 0.9\, \lambda_{max}$; right-top triangle in Fig. 2B), also suggesting that the network strands were fully stretched. This correspondence in the microscopic structures strongly supports the validity of the microscopic description predicting $\lambda_{max}$ in **Eq. 2**. This ideal stretchability was not observed in tetra-branched polymer gels without the occurrence of SIC.

**Table 1. Crystal lattice parameters.** Measured d-spacing ($d_m$) and calculated d-spacing ($d_{hkl}$) values, and the crystallinity ($X$) of a tri-branched gel with a polymer concentration of 240 g/L.

| $q$ (Å$^{-1}$) | $d_m$ (Å) | Reflection surface | $d_{hkl}$ (Å) | Type of crystal | $X$ (%) |
|---|---|---|---|---|---|
| 1.5 | 4.3 | Triclinic (100) | 4.4 | planar zig-zag | 81 |
| 1.8 | 3.6 | Triclinic (010) | 3.7 | planar zig-zag | |

Notably, the SIC observed in the tri-branched gel is reversible. The characteristic diffraction pattern disappeared after removing the stress. In a cyclic loading test ($\lambda = 18 \approx 0.8\lambda_{max}$), almost-constant stress–elongation curves were measured for the tri-branched gel, with slight mechanical hysteresis loss observed at the second and subsequent loadings (Fig. 2D). The slight mechanical hysteresis may be because of a small amount of residual crystalline material that did not dissolve on the timescale of the loading cycle. The unvarying stress–elongation relationship distinguishes tri-branched gels from self-healing gels, which undergo massive mechanical hysteresis (15). SIC, which induces strain-hardening, is considered an on-demand reinforcement devise in polymer gels (34) and vulcanized natural rubber (36). When a stress concentration occurs within a meso-scale region, SIC occurs and hardens the region, inhibiting further stretching and stress concentration at the region. It should be noted that SIC does not increase $\lambda_{max}$, however, it moves $\lambda_b$ closer to $\lambda_{max}$. Such reinforcement by strain-hardening is also observed in metals such as transformation-induced plasticity (TRIP) steel and ceramics showing strain-induced martensitic transformation (37).

Our observations indicate the possibility that polymer networks with reduced branching factors can be efficiently reinforced by reversible SIC. The reversible SIC was observed in tri-branched gels but not tetra-branched gels, in the concentration range tested. Recently, Mayumi and colleagues (34) observed similar reversible SIC in a polymer gel with slidable crosslinks (slide-ring gel) that allow the constituent polymer chain to effectively orient along the stretching axis without fracture, resulting in reversible SIC. On the basis of the current observations, at least two requirements for reversible SIC were hypothesized. First, appropriate stability of the crystal of backbone polymer in the solvent is necessary. Both tri-branched and slide-ring gels contain water and PEG. Thus, the water–PEG system is an essential condition for the current reversible SIC. The PEG crystal is unstable without orientation, but stable under a certain degree of orientation, realizing reversible SIC. This hypothesis is also supported by the observation of reversible flow-induced crystallization in water–PEG systems (38). The PEG concentration range of 200 to 400 g/L is expected to be appropriate for reversible SIC. PEG crystals were not observed at a PEG concentration below 200 g/L. In addition, the PEG crystal would not be reversible at a high PEG concentration (more than 400 g/L), as evidenced by the large residual strain observed in Fig. S1. Notably, no reversible SIC is observed in water–polyvinyl alcohol gels (39), another prominent water-soluble crystalline polymer. The crystal of polyvinyl alcohol is too stable in water to show reversible SIC, and acts as a stable crosslink.

The second requirement is a high chain orientation. This requirement is self-evident because the chain orientation is the driving force of SIC. Significant chain orientation is achieved in slide-ring gels due to the motion of slidable crosslinks (34). Therefore, our observation indicates that the polymer chain effectively orientates in the tri-branched rather than the tetra-branched network. Our hypothesis is supported by the mathematical theory of structural rigidity proposed by Maxwell (40, 41). The model treats networks as a collection of rigid rods with $f$-branched flexible links (Fig. 4). The difference between rigid rods and flexible polymer chains is considered insignificant as the focus is on the chain orientation in branched structures. Figure 4 (A to C) shows 2D schematics describing the deformation of 3-, 4-, and 8-branched networks, respectively. The blue dashed lines represent the flexible axes, along which the network can collapse; the network collapses more easily in the presence of a higher number of flexible axes. The 3-branched network has more flexible axes than the 4-branched network, while the 8-branched network does not have any such axes. According to the model, the network structure does not collapse if the following conditions are satisfied: $f >$

4 in 2D space and $f > 6$ in 3D space. This model clearly shows that networks with smaller $f$ values easily collapse, and the constitutive chains effectively orientate along with the macroscopic deformation. Therefore, 3-branched polymer networks with the smallest $f$, which allows network structure formation, achieve the most effective orientation in the stress-concentrated region at the meso-scale.

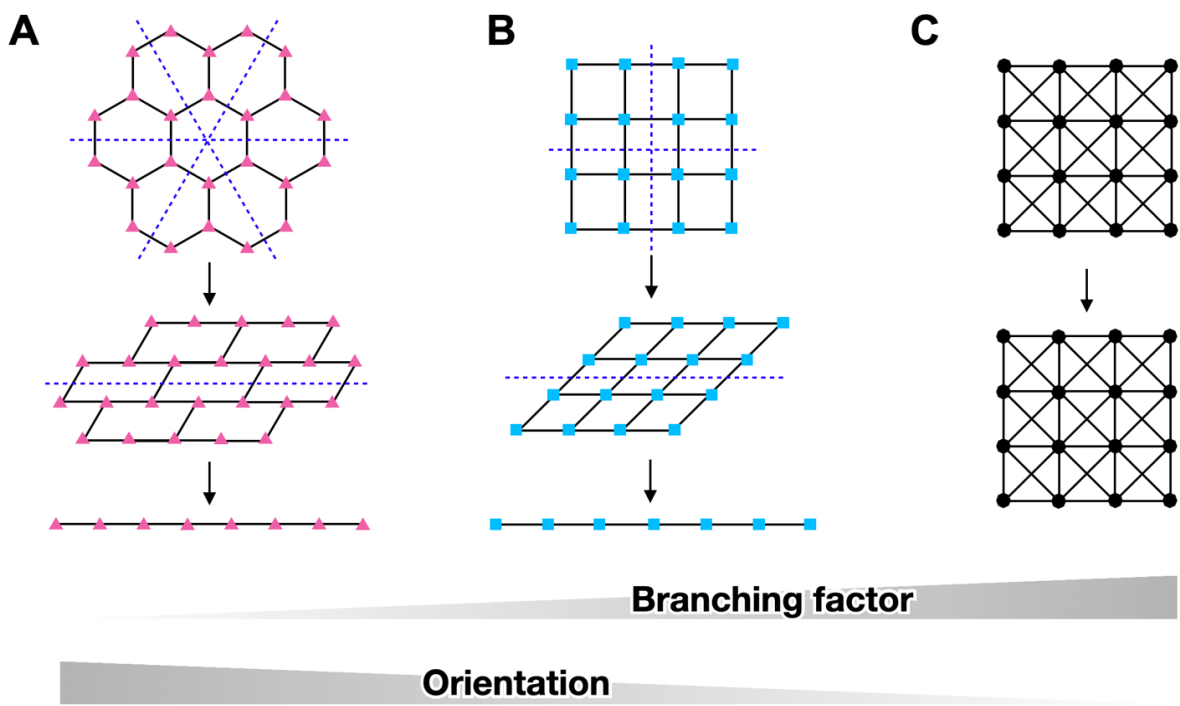

**Fig. 4. Structural rigidity model.** Schematics showing the rigidities of 2D $f$-branched bonded networks ($f = 3$ (**A**, triangles), 4 (**B**, squares), and 8 (**C**, octagons)). The blue dashed lines represent the flexible axes. The networks with $f = 3$ and 4 can be collapsed along the flexible axes, while the network with $f = 8$ cannot be collapsed.

**Discussion**

This study experimentally demonstrated that a polymer network with the lowest branching factor is effectively toughened by reversible SIC. The reversible SIC observed in tri-branched gels acts as an on-demand reinforcement scheme and contributes to the robustness and high strength of these gels. A critical issue that was not assessed in detail is the balance between the stabilities of amorphous and crystalline phases. To realize reversible SIC, appropriate polymer and solvent systems should be selected, and the polymer concentration should be appropriately controlled. Molecular designs restricting the gel swelling will partly contribute to the realization of reversible SIC in practical rubbery materials (42). This work reveals the importance of tuning the branching factor, which is expected to facilitate further advancements in polymer science and help solve some of the global problems caused by rubbery materials.

**Materials and Methods:**

**Fabrication of the gels**

Tri-amine-terminated poly(ethylene glycol) (tri-PEG-PA), tri-NHS-glutarate-terminated poly(ethylene glycol) (tri-PEG-HS), tetra-amine-terminated poly(ethylene glycol) (tetra-PEG-PA), and tetra-NHS-glutarate-terminated poly(ethylene glycol) (tetra-PEG-HS) were purchased from NOF Co. (Tokyo, Japan). The weight-average molecular weights ($M$ g/mol) of the tri-branched prepolymers were 10k, 20k, and 40k, while the tetra-branched prepolymers had $M$ values of 5k, 10k, 20k, and 40k. The molecular weights were measured by gel permeation chromatography with linear PEG as the standard. All prepolymers were used without further purification and were first dissolved in phosphate (PB) or phosphate-citric acid (CPB) aqueous buffers. Previous studies have reported that the interaction parameter between PEG and water ($\chi$) ranges between 0.43 and 0.45 (43, 44). To control the reaction rate, the optimal ionic strength and pH were used (Table 2). To fabricate the tri- and tetra-branched gels, equal amounts of the two corresponding prepolymer solutions were mixed without incubation, and the resulting solution was poured into a dumbbell-shaped mold (JIS K6251 No. 7). As for the tetra-branched gel with $M = 20$k and polymer concentration ($c$) of 140 g/L and tri-branched gel with $M = 10$k and $c$ of 140 g/L, the solution of tetra-PEG-HS was incubated at room temperature (approximately 298 K) during 0 or 3 hours to partially deactivate NHS ester. Herein, the polymer concentration is defined as the prepolymer weight divided by the solution volume, rather than by the solvent volume. The elastic modulus of the 3-h-incubated tetra-branched gel was set to achieve a value similar to that of the tri-branched gel with $M = 20$k and $c = 140$ g/L due to the deactivation of the terminal groups (22). The elastic modulus of the tri-branched gel incubated for 3 hours was much smaller than that of the tri-branched gel incubated for 0 hours (Fig. S4). At least 12 hours was allowed for the reaction to reach completion before the following experiments were performed.

To fabricate tri-tetra-branched gels, tri-PEG and tetra-PEG prepolymers with $M = 20$k were dissolved in PB or CPB (Table 2) to give different molar fractions of tri-branched prepolymers ($s$), where $s$ is defined as follows.

$$s = \frac{[\text{tri} - \text{PEGs}]}{[\text{tri} - \text{PEGs}] + [\text{tetra} - \text{PEGs}]} \quad (4)$$

Tri-tetra-branched gels were formed for $0 < s < 1$, while $s = 0$ and 1 correspond to the pure tetra-branched and tri-branched gels, respectively. Gels with $s$ values of 0, 0.25, 0.5, 0.75, and 1 were prepared. In the tri-tetra-branched gels, tetra-branched and tri-branched prepolymers were selected and mixed so that the number of mutually reactive groups matched at the determined $s$. The polymer solutions were molded and reacted using the process used for the other sample types. Here, the degree of polymerization between the branch points ($N$) was calculated as

$$N = \frac{M}{44}\frac{2}{f} \quad (5)$$

where $f$ is the branching factor of the prepolymer. The $N$ values of the tri-tetra-branched gels were calculated as the number average:

$$N = 303s + 227(1 - s) \quad (6)$$

where 303 and 227 correspond to the $N$ of tri-PEG and tetra-PEG prepolymers, respectively, calculated using **Eq. 5**.

**Table 2. Fabrication conditions of tri-branched, tri-tetra-branched, and tetra-branched gels.** Overlap concentrations $c^*$ of tetra-branched prepolymers were taken from Ref. 26.

| Type of gel | $s$ (-) | $M$ (g/mol) | $N$ (-) | $c$ (g/L) | $c^*$ (g/L) | Buffer for PA | Buffer for HS |
|---|---|---|---|---|---|---|---|
| Tri-branched gel | 1 | 10k | 150 | 75 | 61 | PB pH 7.4 25 mM | CPB pH 5.8 25 mM |
| | | | | 140 | | PB pH 7.4 100 mM | CPB pH 5.8 100 mM |
| | | 20k | 300 | 75 | | PB pH 8.0 100 mM | PB pH 7.4 100 mM |
| | | | | 140 | | PB pH 7.4 100 mM | PB pH 7.4 100 mM |
| | | | | 240 | | PB pH 7.4 100 mM | CPB pH 5.8 100 mM |
| | | 40k | 610 | 140 | Not measured | PB pH 8.0 100 mM | PB pH 7.4 100 mM |
| Tri-tetra-branched gel | 0.75 | 20k | 280 | 140 | | PB pH 7.4 100 mM | CPB pH 5.8 100 mM |
| | 0.5 | | 270 | | | PB pH 7.4 100 mM | CPB pH 5.8 100 mM |
| | 0.25 | | 250 | | | PB pH 7.4 100 mM | CPB pH 5.8 100 mM |
| Tetra-branched gel | 0 | 5k | 57 | 140 | 110 [Ref. 26] | CPB pH 5.8 100 mM | CPB pH 5.8 100 mM |
| | | 10k | 110 | | 70 [Ref. 26] | PB pH 7.4 100 mM | CPB pH 5.8 100 mM |
| | | 20k | 230 | | 40 [Ref. 26] | PB pH 7.4 100 mM | CPB pH 5.8 100 mM |
| | | 40k | 460 | | 15 [Ref. 26] | PB pH 7.4 50 mM | CPB pH 5.8 50 mM |

Notably, the above protocol estimating $N$ is different from the conventional one using the elastic modulus ($G$). Classical models of rubber elasticity predicts $G$ using $G = 2nk_BT$ (45) and $G = (1-2/f)nk_BT$ (8) in affine network and phantom network models, respectively. Here, $k_B$ is the Boltzmann constant and $T$ is the absolute temperature. These predictions enable the estimation of the number density of prepolymers ($n$) from $G$, and subsequent estimation of $N$ from $n$, as the scaling relationship $G \sim n \sim c/N$. A combination of the scaling relationship and Kuhn's model (Eq. 11-2) results in the scaling prediction $\lambda_{Kuhn} \sim G^{-1/2}$. However, a previous study (22) revealed limitations when using Kuhn's model to describe tetra-branched gels. This model cannot predict a series of results as a function of the network connectivity ($p$), which is the fraction of the reacted terminal functional groups to all such groups. The increase in $p$ leads to an increase in $G$ but does not influence the stretch ratio at break ($\lambda_b$) (Fig. S4), in contrast to the prediction of Kuhn's model ($\lambda_{Kuhn} \sim G^{-1/2}$). Instead, a series of experimental results of tetra-branched gels showed the scaling relationship shown by **Eq. 1**.

Furthermore, our recent studies indicated that classical models are not suitable for describing polymer gels (23, 24). The elastic moduli of polymer gels are not proportional to $T$ but have a significant negative constant $b$ as $G = \alpha T + b$, where $\alpha$ and $b$ depend on the microscopic network structure. Therefore, the microscopic parameters such as $n$ and $N$ cannot be predicted from the classical models. In contrast, it was possible to correlate the fracture energy of a series of polymer gels and the prediction using $N$, estimated as two times the polymerization degree of the prepolymer arm. The fracture energy was quantitatively explained by the Lake–Thomas model with a constant of the order of 1 (25). On the basis of this correspondence, $N$ was estimated from **Eq. 6**.

**Stretching tests**

Uniaxial tensile tests were conducted using a Shimadzu AG-X Plus universal tester (Shimadzu Corporation, Japan) at room temperature (approximately 298 K). Dumbbell-shaped specimens (JIS K6251 No. 7) were uniaxially stretched at an initial strain rate of ~0.07 s$^{-1}$ (results in Fig. 1) or ~0.16 s$^{-1}$ [all results of Fig. 2 (A and B) and results of tri-branched gels of Fig. S4]. The strain rate was accelerated for the tests giving the data shown in Fig. 2 (A and B) to minimize the effect of water evaporation from the specimen. The stress–elongation ratio ($\sigma$–$\lambda$) curves were automatically recorded by the universal tester. Uniaxial tensile tests were conducted at least three times for each sample type, and typical results are shown in Fig. 2A. The average $\lambda$ at rupture and its standard deviation were used as the representative $\lambda_b$ and the experimental error shown in Fig. 2B, respectively.

Loading–unloading tests for the tri-branched gel (results in Fig. 2D) were conducted using a Sun Rheometer 150ST and CR-500DX system (Sun Scientific Corporation, Japan). Rectangular samples (5 mm wide, 1 mm thick, and approximately 3.7 mm long) were stretched to $\lambda \approx 18$ and then released to return to their original length at an initial strain rate of ~0.3 s$^{-1}$. The $\sigma$–$\lambda$ curves were automatically recorded by the tester. The experiments were conducted in a custom-designed liquid paraffin chamber to prevent solvent evaporation at 25 °C.

**Maximum stretch ratio**

To derive **Eq. 3**, it was assumed that the center of the prepolymers (before the gelation reaction is initiated) has a random close-packed arrangement, and that its placement determines the polymer network structure. Under these assumptions, the linear distance between the branch points of the polymer network ($d$) was estimated as the diameter of a sphere, which distributes with the number density of prepolymers ($n$) and the packing density ($\Phi$). Therefore

$$\Phi = \frac{4}{3}\pi \left(\frac{d}{2}\right)^3 n \quad (7)$$

which is rewritten as

$$d = \left(\frac{6\Phi}{\pi n}\right)^{1/3} \quad (8)$$

Because the number density of prepolymers satisfies $n = cN_A/(m_{mono}N)$, the maximum stretch ratio $\lambda_{max} \equiv aN/d$ is calculated as

$$\lambda_{max} = aN\left(\frac{\pi n}{6\Phi}\right)^{1/3} = a\left(\frac{\pi N_A}{6\Phi m_{mono}}\right)^{1/3} c^{1/3} N^{2/3} \quad (9)$$

which is identical to **Eq. 2** with **Eq. 3**.

**Failure of Kuhn's model for fracture**

Kuhn's model for fracture predicts the maximum stretch ratio of rubbery materials by considering a replica of the network strand in the gel considered. The chemical structure and $N$ of the replica polymer chain are identical to those of the considered network strand. It exists in a polymer solution with the same polymer volume fraction $\phi$ as the considered gel. The maximum stretch ratio ($\lambda_{Kuhn}$) is the end-to-end distance ratio of the replica in the fully stretched condition to that in the initial condition ($d_{Kuhn}$).

$$\lambda_{kuhn} = \frac{aN}{d_{Kuhn}} \quad (10)$$

When the solvent is good for the polymer chain, such as the PEG–water system, $d_{Kuhn}$ depends on $\phi$ and obeys the following equations (31).

$$d_{Kuhn} = R_F = a_0 \left(\frac{v_0}{a_0^3}\right)^{2\nu-1} N_0^\nu \quad \text{(for } \phi < \phi^*\text{)} \quad (11-1)$$

$$d_{Kuhn} = R_F \left(\frac{\phi}{\phi^*}\right)^{\frac{2\nu-1}{2(1-3\nu)}} \quad \text{(for } \phi^* < \phi < \phi^{**}\text{)} \quad (11-2)$$

$$d_{Kuhn} = a_0 N_0^{1/2} \quad \text{(for } \phi^{**} < \phi\text{)} \quad (11-3)$$

Here, $R_F$ is the end-to-end distance in dilute conditions, $\nu$ is the scaling exponent, $a_0$ is the Kuhn length, and $N_0$ is the number of the Kuhn segment, holding a conservation law $a_0 N_0 = aN$. Approximately three monomeric units form the Kuhn segment in the PEG–water system: $a_0 \simeq 3a \simeq 1.0$ nm, and $N_0 \simeq N/3$ (27). Further, $v_0$ is the second virial coefficient and $\phi^*$ is the polymer volume fraction where chain overlaps occur. $\phi^{**}$ is the polymer volume fraction where chains are ideal on all length scales, and the characteristic parameter for each polymer-solvent system is approximately 1 (31). Given that $\phi^* < 0.1$ (Table 2), the experimental condition examined in this study is safely in the range of **Eq. 11-2**.

When the maximum stretch ratio is calculated on the basis of **Eqs. 10** and **11-2**, $\lambda_{Kuhn}=$ 8.1 for the tri-branched gel ($M = 20$ kg/mol, $c = 240$ g/L, $\phi = 0.21$). The similarity between tetra-branched and tri-branched polymer (46) is assumed, and $\nu \simeq 0.59$ and $\phi^* = 0.04$ are accepted on the basis of previous studies on tetra-branched gels (26, 47). Moreover, it was assumed that $(v/a_0^3)^{2\nu-1} \approx 1$ (athermal solvent). This value of $\lambda_{Kuhn}$ is smaller than $\lambda_b$ and the elastic range shown in Fig. 2D. If the prediction is valid, then the tri-branched gel would show an elastic behavior beyond the theoretical limit of stretch ratio. Notably, when **Eq. 11-3** is used, which predicts the highest limit of $\lambda_{Kuhn}$, $\lambda_{Kuhn} = 10 < \lambda_b$. Therefore, Kuhn's model predicts a $\lambda_{Kuhn}$ lower than $\lambda_b$ regardless of the chain conformation. The same is observed in other tri-branched gels ($M = 20$ kg/mol, $c = 140$ g/L, $\phi = 0.12$). In general, elastic materials break at a stretch ratio below the theoretical limit, i.e., $\lambda_{max} > \lambda_b$, due to the stress concentration (37, 48). Therefore, by respecting the general relationship, Kuhn's model shows a significant underestimation of $\lambda_{max}$ in tri-branched gels.

It is expected that Kuhn's model for fracture seems to be suitable for describing network formation in dense polymer solutions. A reactive end can find other reactive ends nearby, and each chain can form a network structure while maintaining the probable end-to-end distance. Therefore, the end-to-end distance of the chains is not affected by the networking reaction, and the probable end-to-end distance is appropriate as the reference length. In contrast, the situation changes drastically for network formation in dilute branched prepolymer systems forming polymer gels. It is difficult to find reactive ends within the range of the probable end-to-end distances. Therefore, a reactive end binds to another end within the range where the end can fluctuate, and the end-to-end distance is inevitably affected by the reaction. Notably, these reactions proceed without incurring a large penalty, because the energy required to change the end-to-end distance by the probable end-to-end distance is at most $k_BT$ (49). The high reaction yields of up to 90%, even in dilute systems (26), strongly support this microscopic scheme. Therefore, in a dilute system, the distance between the ends is determined by the density of the branch points, corresponding to the treatment of **Eq. 9**. Our experimental observation suggests the microscopic picture of Fig. 2B holds even in the semi-dilute region, because $\lambda_b$ obeys **Eq. 1** up to approximately 10 times the overlapping polymer concentration $c^*$.

### *In-situ* small/wide angle X-ray scattering

*In-situ* SAXS/WAXS measurements were conducted at the BL05XU beamline of the SPring-8 synchrotron, Japan. An X-ray wavelength of 1.0 Å was used, and the sample-to-detector distances were 3950 mm and 300 mm, respectively, for SAXS and WAXS tests. The samples were exposed to X-ray irradiation for 1 s while being stretched using a custom tensile-testing apparatus. The stretching rate was ~0.3 s$^{-1}$. The scattering patterns were obtained using a 2D detector (Pilatus 1M; Dectris, Switzerland). The transmittance was calculated from the ratio of the incident beam intensity, measured using an ionization chamber, to the transmitted beam intensity, measured using a PIN diode embedded in a beam stopper.

The 2D WAXS patterns were converted to 1D intensity profiles from the direction perpendicular to stretching by symmetric sector averaging over azimuthal angles of 90 ± 2.5°. The average scattering intensity was normalized by the transmission intensity and sample thickness and then plotted against the amplitude of the scattering vector ($q$)

$$q = \frac{4\pi}{\lambda_X} \sin\left(\frac{\theta}{2}\right) \quad (12)$$

where $\lambda_X$ is the X-ray wavelength and $\theta$ is the scattering angle calibrated from the diffraction pattern of silver behenate. The obtained 1D WAXS profiles were fitted with Gaussian functions and were decomposed into a number of constituent peaks (Fig. S3). Characteristic WAXS diffraction peaks were indexed to the reflections from the crystal lattice, based on the agreement between the measured and the calculated d-spacing, $d_m$ and $d_{hkl}$, respectively (Table 1). The values of $d_m$ and $d_{hkl}$ were calculated as follows:

$$d_m = \frac{2\pi}{q} \quad (13)$$

$$d_{hkl} = \frac{2\pi}{|h\mathbf{a}^* + k\mathbf{b}^* + l\mathbf{c}^*|} \quad (14)$$

where $h$, $k$, and $l$ are the Miller indices, and $\mathbf{a}^*$, $\mathbf{b}^*$, and $\mathbf{c}^*$ are the basic translation vectors of the reciprocal lattice, which were calculated from the crystal lattice structures (11). The crystallinity $X$ was calculated as follows:

$$X = \frac{A_c}{A_c + A_a} \quad (15)$$

where $A_c$ is the integrated area of the peaks assigned to the crystalline phase, and $A_a$ is that of the peaks assigned to the amorphous phase.

**Dynamic viscoelasticity measurement**

Dynamic viscoelasticity measurements were conducted to estimate the overlapping polymer concentration $c^*$ of tri-maleimide-terminated poly(ethylene glycol) (tri-PEG-MA) and tri-thiol-terminated poly(ethylene glycol) (tri-PEG-SH) with $M$ = 10k, purchased from NOF Co. (Tokyo, Japan). Tri-PEG-MA and tri-PEG-SH were dissolved in CPB (pH 3.8, 200 mM) by tuning $c$ from 0 to 110 g/L. The shear viscosities ($\eta$) of the solutions were measured using the cone-plate geometry of a rheometer (MCR302; Anton Paar, Graz, Austria) at a constant shear rate of 100 s$^{-1}$ at 25 °C. The specific viscosity ($\eta_{sp}$) was calculated as

$$\eta_{sp} = \frac{\eta - \eta_0}{\eta_0} \quad (16)$$

where $\eta_0$ is the shear viscosity of the solvent. The results indicated a crossover from dilute scaling ($\eta_{sp} \sim c$) to semi-dilute scaling ($\eta_{sp} \sim c^2$), as shown in Fig. S5 (50, 51). $c^*$ was estimated at the intersection point of the two scaling laws. Further, $c^*$ is independent of the end-groups, and the average values of $c^*$ are shown in Table 2.

**Acknowledgements:**

**Funding:** This work was supported by Society for the Promotion of Science (JSPS) Grant-in-Aid for JSPS Research Fellows awarded to T.F. (number 19J22561); Early Career Scientist Grant awarded to N.S (number 19K14672); Scientific Research (A) Grant awarded to T.S (number 21H04688); Transformative Research Area Grant awarded to T.S (number 20H05733); Japan Science and Technology Agency (JST) CREST grant awarded to T.S (number JPMJCR1992).

**Author contributions:** Conceptualization: T.S. Methodology: T.F., N.S., C.L., K.M., and T.S. Investigation: T.F., C.L., and K.M. Visualization: T.F., N.S., C.L., K.M., and T.S. Funding acquisition: T.F., N.S., U.C., and T.S. Project administration: U.C. and T.S. Supervision: T.S. Writing - original draft: T.F. and T.S. Writing - review & editing: N.S., C.L., T.K., and U.C.

**Competing interests:** The authors declare no competing interests.

**Data Availability:** All data needed to evaluate the conclusions in the paper are present in the paper and/or the Supplementary Materials.

# Supplementary Materials

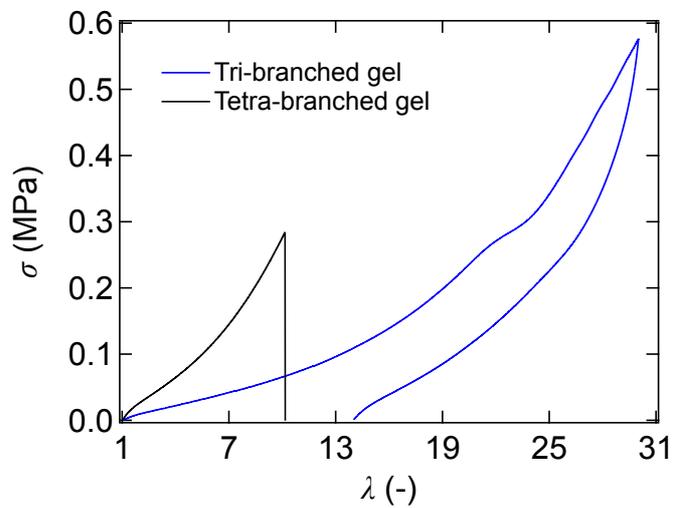

**Fig. S1. Typical $\sigma$–$\lambda$ curve of a tri-branched gel (blue) and a tetra-branched gel (black) at a low strain rate.** Here, the polymer concentration was 140 g/L, and the molecular weight of the prepolymers was 40k. The strong residual strain in the tri-branched gel was due to the crystallization caused by water evaporation.

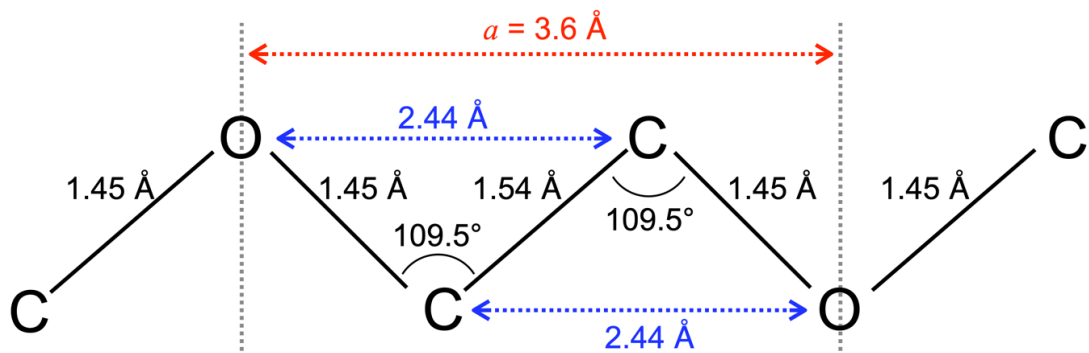

**Fig. S2. Schematic illustration of the effective length of monomer segment ($a$ = 3.6 Å).**

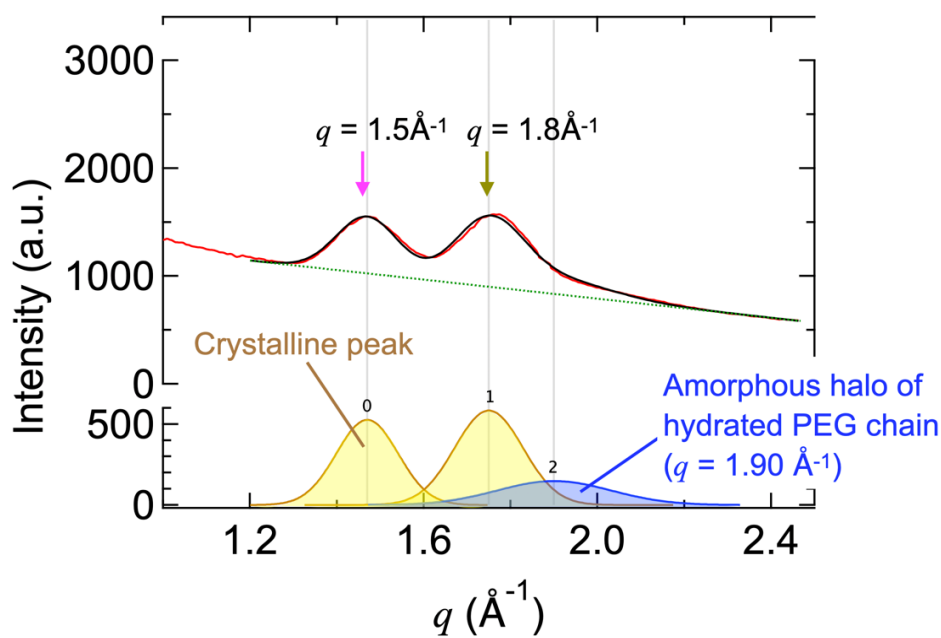

**Fig. S3. 1D WAXS profile of a tri-branched gel with a polymer concentration of 240 g/L at $\lambda = 18$.** The red, black, yellow, and blue lines represent the raw data, the best fitting result, the crystalline peaks, and the amorphous halo, respectively. The green dotted line represents the baseline.

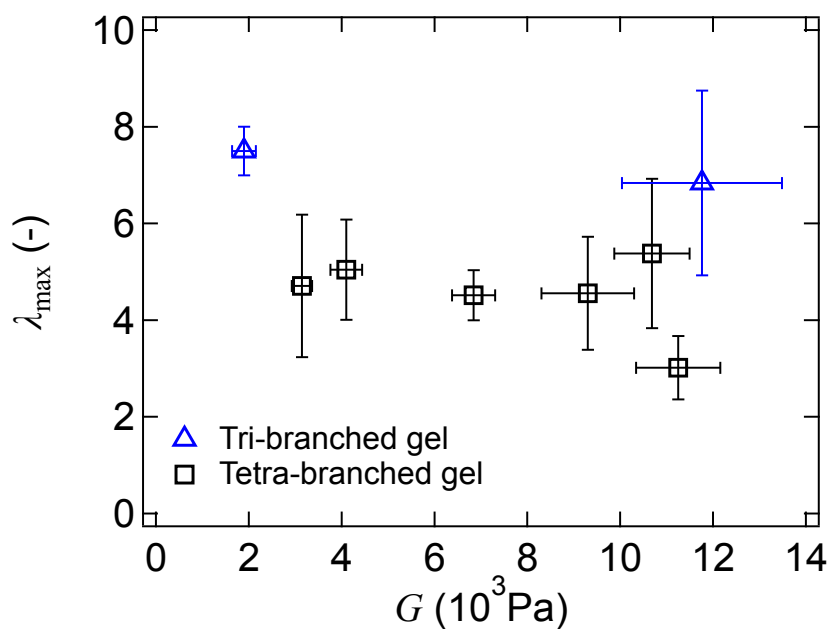

**Fig. S4. Relationship between $\lambda_{max}$ and $G$ of tri-branched gels with $M = 10k$ and $c = 140$ g/L (blue triangles) and tetra-branched gels with $M = 20k$ and $c = 92$ g/L (black squares).** Here, we tuned $G$ at constant $M$ and $c$ by incubating tetra-PEG-HS solutions and deactivating the end-groups of tetra-PEG-HS. In contrast to Kuhn's model ($\lambda_{max} \sim G^{-1/2}$), $\lambda_{max}$ is independent of $G$. The data were reproduced from our previous study (22).

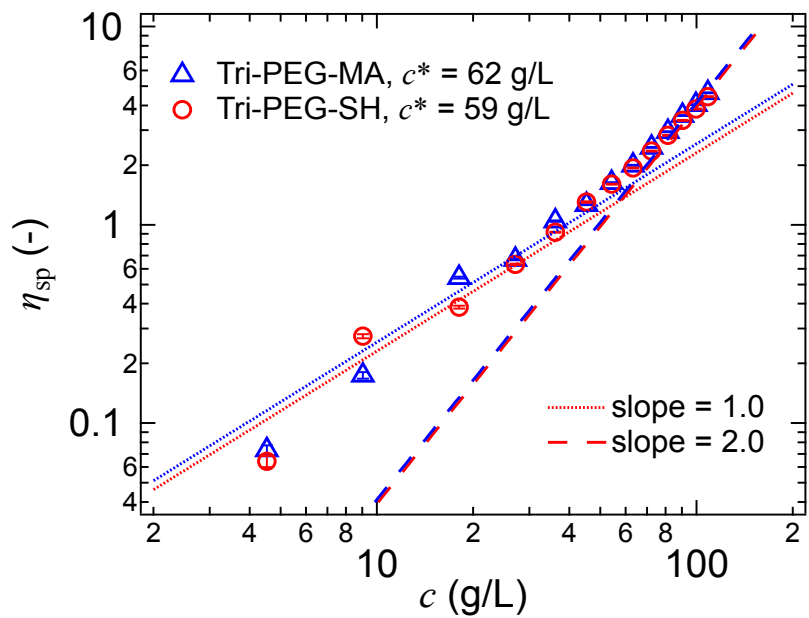

**Fig. S5. Relationship between $\eta_{sp}$ and $c$ of tri-PEG-MA (blue triangles) and tri-PEG-SH (red circles) with $M$ = 10k.** We found a crossover from dilute scaling ($\eta_{sp} \sim c$, dotted lines) to semi-dilute scaling ($\eta_{sp} \sim c^2$, dashed lines). (44, 45) We estimated $c^*$ at the intersection point of the two scaling laws and confirmed the end-group independent $c^*$.